\documentclass[twocolumn]{aastex63}
\usepackage{amsmath}
\usepackage{mathtools}
\usepackage{amssymb}
\usepackage{lineno}

\shorttitle{FRBs' Spectra}
\shortauthors{Zhong et al.}

\begin{document}
\title{Observed Steep and Shallow Spectra, Narrow and Broadband Spectra, Multi-frequency Simultaneous Spectra, and Statistical Fringe Spectra in Fast Radio Bursts: Various Faces of Intrinsic Quasi-periodic Spectra?}
\author[0000-0002-1766-6947]{Shu-Qing Zhong}
\affil{School of Science, Guangxi University of Science and Technology, Liuzhou 545006, People’s Republic of China; sq\_zhong@qq.com}
\author[0000-0003-0055-5287]{Wen-Jin Xie}
\affil{CAS Key Laboratory of Space Astronomy and Technology, National Astronomical Observatories, Chinese Academy of Sciences, Beijing 100012, People's Republic of China}
\author{Jia-Hong Gu}
\affil{School of Science, Guangxi University of Science and Technology, Liuzhou 545006, People’s Republic of China; sq\_zhong@qq.com}

\begin{abstract}
In this paper, through analysis, modelings, and simulations, we show that if the spectra of fast radio bursts (FRBs) are intrinsically quasi-periodic spectra, likely produced by coherent curvature radiation from quasi-periodic structured bunches, then the observed steep and shallow spectra, narrow and broadband spectra, multi-frequency simultaneous spectra, as well as possible statistical fringe spectra in FRBs, could all be various manifestations of these intrinsically quasi-periodic spectra. If so, the period properties of the structured bunches, as inferred from the observed multi-frequency simultaneous spectra and potential statistical fringe spectra, may provide valuable insights into the mechanisms behind the formation of such structured bunches.
\end{abstract}


\section{Introduction}
\label{sec:introduction}
The spectra of fast radio bursts (FRBs) are crucial for exploring their radiation mechanism(s) \citep[see reviews][and references therein]{cor19,pet19,xiao21,zhang23}. Observationally and statistically, FRB spectra exhibit significant diversity, e.g., steep and shallow spectra \citep{spi16,mac19,chime21}, narrow and broadband spectra \citep{kumar21,zhou22,zhangyk23,kumar24,ple21}, 
multi-frequency simultaneous spectra \citep{law17,chime20,boc20}, and statistical fringe spectra \citep{lyu22,lyu23}.

The observed spectral shape of FRBs is often approximated by a power-law function $F_{\nu}\propto\nu^{-\alpha}$. Many FRB bursts exhibit shallow spectra with indices $|\alpha|<3$ \citep{mac19,chime21,zhong22}, while others, both from non-repeating and repeating FRBs, display steep spectra with indices $|\alpha|>3$. For example, FRB 20110523A had $\alpha=7.8\pm0.4$ \citep{masui15}, 
FRB 20121102A's bursts spanned a range of $\alpha=-10.4$ to $+13.6$ \citep{spi16}, 
and many cases exceeded $\alpha>10$ in the first CHIME/FRB FRB catalog \citep{chime21}. These steep spectra may also suggest a narrow bandwidth. This is because, if the intrinsic spectra are described by a broken power-law and the observed spectra capture only the rising or falling segments described by a single power-law, the relationship between spectral index and bandwidth, as presented in the equation (2) of \cite{yang23}, shows that increasing the absolute value of either the rising index $\alpha_l$ or the falling index $\alpha_h$ narrows the bandwidth for a given full width $N_{\rm FW}$. 

Although broadband spectra are commonly observed in FRBs, 
extremely narrow spectra have also been reported in some FRBs, such as FRBs 20190711A, 20201124A, 20220912A, 20240114A \citep{kumar21,zhou22,zhangyk23,kumar24}. 
While these narrow spectra could arise from intrinsic radiation mechanisms or interference processes\footnote{The radiative transfer processes in circumburst medium or emission region (magnetosphere), such as absorption and scattering effects, do not appear to significantly alter the observed spectra or produce narrow/steep spectra \citep{yang23,wang24}.}, 
they seem to be more likely produced by a coherent process, e.g., bunching or maser mechanism \citep{yang23,liu23,wang24}. Interference processes like scintillation, gravitational lensing, or plasma lensing require specific conditions or further observational validation. As discussed and concluded in \cite{yang23}, for instance, scintillation with $\Delta \nu_{\rm sci}\gtrsim100$ MHz demands an intermediately dense, turbulent plasma screen at $\sim10^{15}$ cm from the FRB source. Gravitational lensing cannot explain the observed burst-to-burst variation of the spectra, although it can generate narrow spectra via planet-like objects. While for the plasma lensing, the typical lensing timescale of decades of seconds can cause the spectrum variation during this short time. This requires to be verified by more FRBs in the future.

Multi-frequency simultaneous spectra have been observed only in FRBs 20121102A \citep{law17} and 20200428D\footnote{It is previous FRB 20200428A, now is referred to as FRB 20200428D, as suggested by \cite{giri23}, since three earlier bursts were found.
In addition to FRBs 20121102A and 20200428D, a burst from FRB 20180916B was also simultaneously observed by the Robert C. Byrd Green Bank Telescope (GBT; 300-400 MHz) and CHIME (400-600 MHz), see \cite{cha20}. However, the simultaneous detection in adjacent frequency bands is likely due to downward drifting \citep{pell24}.} 
currently \citep{chime20,boc20}. 
One burst in FRB 20121102A was simultaneously detected at $\sim1.4$ GHz (Arecibo telescope) and $\sim3$ GHz (VLA telescope), see \cite{law17}. 
Since the power-law spectral index 
derived from observations at $\sim1.4$ GHz and $\sim3$ GHz was not in agreement with 
the spectral index limit from the non-detection at 4.85 GHz conducted simultaneously by the Effelsberg telescope, 
\cite{law17} concluded that the broadband burst spectrum could not be described by a single power-law.

Statistical spectral fringe patterns in FRBs 20121102A and 20190520B have emerged based on peak frequency distributions\footnote{In the absence of peak frequency data for bursts, \cite{lyu22} and \cite{lyu23} approximate the central frequencies as the peak frequencies. We adopt the same approach in this work.} \citep{lyu22,lyu23}, though it is still unclear whether such fringe patterns result from statistical and/or observational selection effects, such as the signal-to-noise ratio threshold, detection algorithm, frequency range, burst bandwidth, and/or statistical fluctuation, as elaborated in \cite{lyu22}. 

These observed spectral characteristics motivate us to think that the intrinsic FRB spectra may be quasi-periodic. If such quasi-periodic spectra stem from a coherent process,
such as coherent curvature radiation by quasi-periodic structured bunches moving along the same trajectory proposed by \cite{yang23}, the observed steep and shallow spectra, 
narrow and broadband spectra, multi-frequency simultaneous spectra, and statistical fringe spectra could all be manifestations of such intrinsic quasi-periodic spectra. 
This paper aims to explore this. 

The structure of the paper is organized as follows. Section \ref{sec:simulations} explores how intrinsic quasi-periodic FRB spectra can result in observed steep, shallow, narrow, broadband spectra, multi-frequency simultaneous spectra, and statistical fringe spectra. Section \ref{sec:summary} provides a summary and discussion.

\begin{figure}
	\includegraphics[scale=0.5]{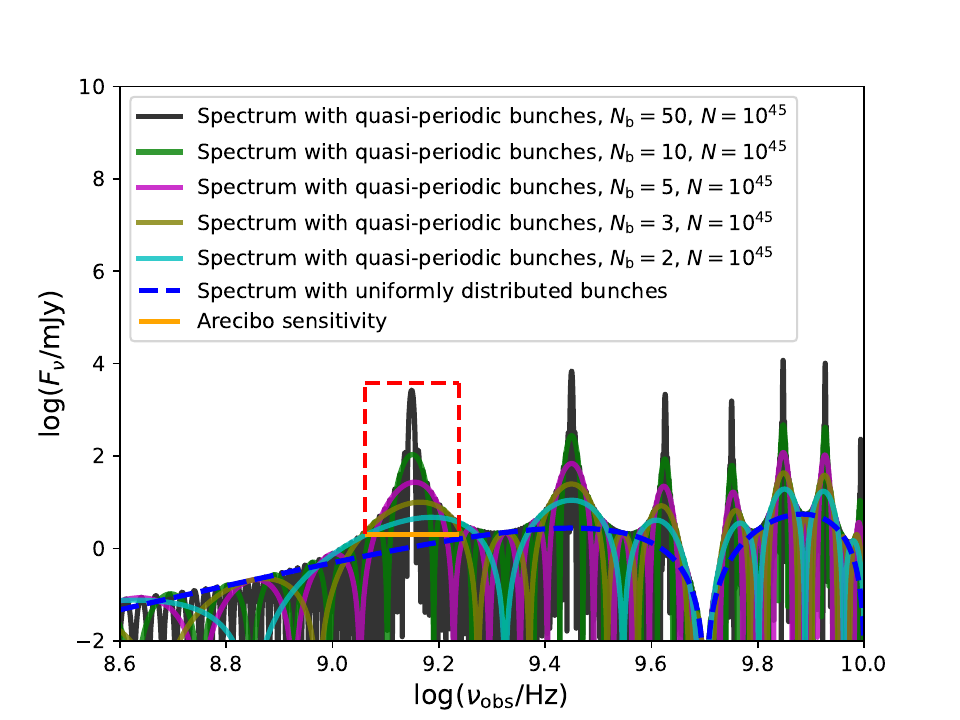}
	\includegraphics[scale=0.5]{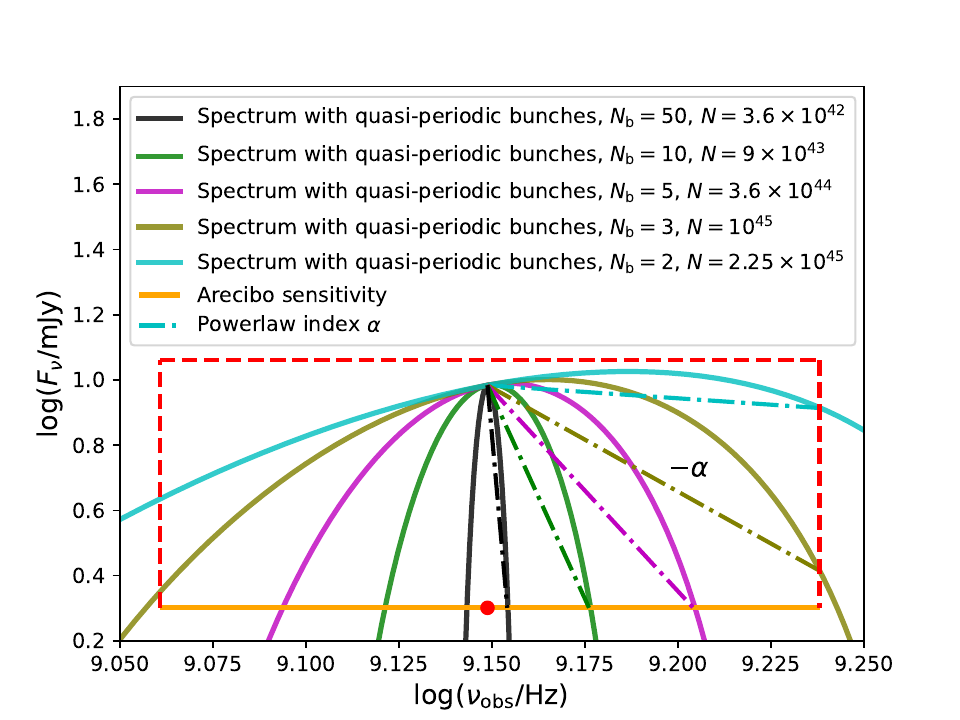}
	\includegraphics[scale=0.5]{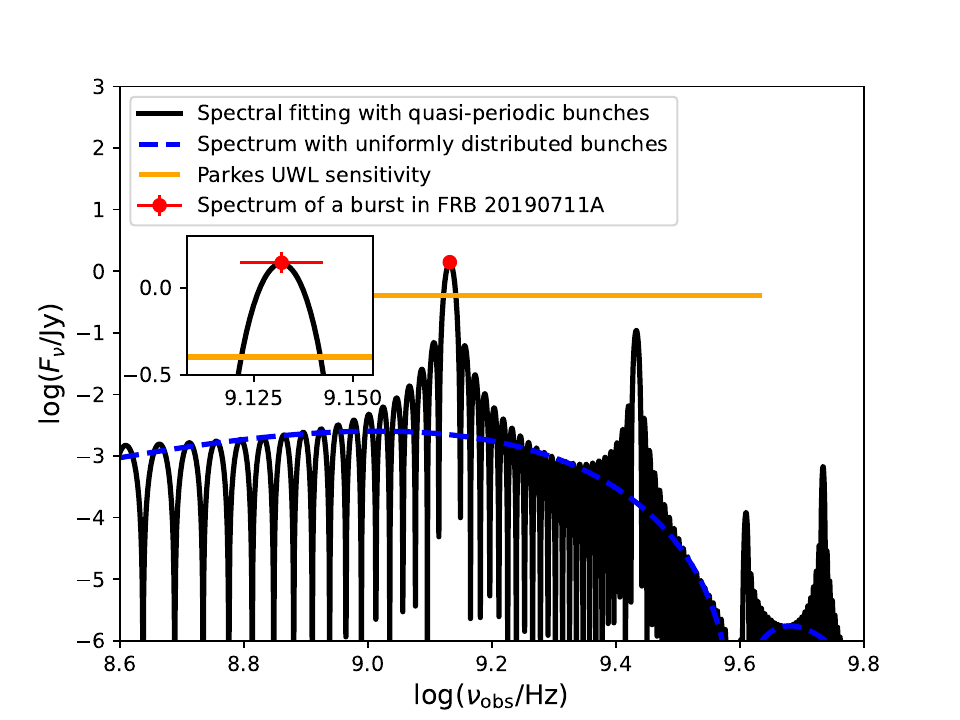}
	\caption{ {\em Top panel}. The spectra produced by coherent curvature radiation from quasi-periodic structured bunches are shown for different bunch numbers per cluster, $N_{\rm b}=2$, 3, 5, 10, and 50 (solid lines), compared to spectra with uniformly distributed bunches which is actually identified with $N_{\rm b}=1$ (blue dashed line). {\em Middle panel}. The spectra above the telescope sensitivity, corresponding to the red dashed rectangle along the first peak in the upper panel, are displayed for similar peak flux values, achieved by only adjusting $N=N_{\rm c}N_{\rm p}^2$. The dashed-dotted lines represent approximate power-law fittings for the falling portion of the spectra. {\em Bottom panel}. A case study: the fitting for the most fractionally narrow-banded burst in FRB 20190711A. The inset is a zoom for the spectral data, fitting line, and sensitivity along the first peak frequency.}
	\label{fig:steep}
\end{figure}

\section{Spectral Analysis, Modelings, and Simulations}
\label{sec:simulations}
For coherent curvature radiation emitted by strictly periodic multiple bunches moving along the same trajectory\footnote{Since coherent radiation is still radiated into multiples of $2\pi\omega_m$ with narrow bandwidths, even for periodically distributed bunches with relative random phase \citep{yang23}, we use strictly periodically distributed bunches in this work for convenience.}, 
the intrinsic spectrum is quasi-periodic and can be expressed as \citep{yang23}
\begin{equation}
	\frac{dI_{\rm c}}{d\omega d\Omega}=\left|E_{N_{\rm p}}(\omega)\right|^2 \sin ^2\left(\frac{N_{\rm b} \omega}{2 \omega_m}\right) \sin ^{-2}\left(\frac{\omega}{2 \omega_m}\right),
	\label{eq:dI_c}
\end{equation}
where $N_{\rm b}$ is the number of bunches, and $P_m=1/\omega_m$ represents the period of the bunch distribution.
The coherent peak frequencies appear at $\omega / \omega_m=2 n \pi$, where $n \in \mathbb{Z}^{+}$. 
The spectrum of a single bunch is given by
\begin{equation}
\left|E_{N_{\rm p}}(\omega)\right|^2= 2 \left[1-\cos\left(\frac{\omega\Delta}{c}\right)\right] N_{\rm p}^2\left|E_1(\omega)\right|^2,
	\label{eq:E_N}
\end{equation}
which assumes an electron–positron pair bunch, where each pair bunch consists of one electron clump and one positron clump separated by a distance $\Delta$ \citep{yang20}. Here, $N_{\rm p}$ is the number of coherent pairs in one bunch, i.e., the number of electron/positron in each clump. The reason for using this type of bunch will be discussed in Section \ref{subsec:multifrequency}.
$\left|E_1(\omega)\right|^2$, the spectrum of curvature radiation by a single charge, is described by \citep{yang18}
\begin{equation}
	\left|E_1(\omega)\right|^2 \simeq \frac{e^2}{c}\left[\frac{\Gamma(2 / 3)}{\pi}\right]^2\left(\frac{3}{4}\right)^{1 / 3}\left(\frac{\omega \rho}{c}\right)^{2 / 3} e^{-\omega / \omega_c},
	\label{eq:E_1}
\end{equation}
where $\omega_c=3 c \gamma^3 /(2 \rho)$ is the characteristic frequency of curvature radiation, with $\rho$ as the curvature radius, 
$\gamma$ the Lorentz factor, and $c$ the speed of light.

It is important to note that the term $\sin^2\left(\frac{N_{\rm b} \omega}{2 \omega_m}\right) \sin^{-2}\left(\frac{\omega}{2 \omega_m}\right)$,
which is related to the coherence properties of multiple bunches, implies a higher level coherence. Not only are the separated electron-positron pairs within each bunch coherent, but also the bunches themselves form coherent ``clusters'' we call. As a result, the total power spectrum from coherent curvature radiation by these bunch-forming clusters is
\begin{equation}
\begin{aligned}
	\frac{dI_{\rm tot}}{d\omega d\Omega}=&N_{\rm c} \frac{dI_{\rm c}}{d\omega d\Omega}  \\
	=&2N_{\rm c} N_{\rm p}^2 \left[1-\cos\left(\frac{\omega\Delta}{c}\right)\right]  \\
	&\times \left|E_1(\omega)\right|^2 \sin^2\left(\frac{N_{\rm b} \omega}{2\omega_m}\right) \sin^{-2}\left(\frac{\omega}{2\omega_m}\right) \\
	=&2N \left[1-\cos\left(\frac{\omega\Delta}{c}\right)\right]  \\
	&\times \left|E_1(\omega)\right|^2 \sin^2\left(\frac{N_{\rm b} \omega}{2 \omega_m}\right) \sin^{-2}\left(\frac{\omega}{2 \omega_m}\right),
\end{aligned}
	\label{eq:dI_tot}
\end{equation}
where $N_{\rm c}$ is the number of clusters, and $N=N_{\rm c}N_{\rm p}^2$ is set. Note that in this case $N_{\rm b}$ would represent the number of bunches in one cluster, rather than the total number of bunches. However, if $N_{\rm b}=1$, which corresponds to the case of uniformly distributed bunches, $N_{\rm c}$ would revert to being the total number of bunches.

Such that the observed flux $F_{\nu}$, representing the energy received per unit time per unit frequency per unit area, is written by \citep{yang18}
\begin{equation}
	F_\nu=\frac{2\pi}{TD^2} \frac{d I_{\rm tot}}{d \omega d \Omega},
	\label{eq:Fv}
\end{equation}
where $D$ is the distance between the source and observer, and $T$ is the time interval between adjacent clusters. For simplicity, we assume $T\sim 1/\nu_c\sim 1{\rm ns}$ for $\nu_c\sim{\rm GHz}$.

\begin{deluxetable*}{ccccc}
	\label{tab:modeling}
	\tablecaption{Parameter Values for Modeling Multi-frequency Simultaneous Spectra and Simulating Statistical Fringe Spectra}
	\tablehead{
		Parameters &
		\multicolumn{2}{c}{Multi-frequency Simultaneous Spectra} &
		\multicolumn{2}{c}{Statistical Fringe Spectra}\\
		&
		FRB 20121102A &
		FRB 20200428D &
		FRB 20121102A &
		FRB 20190520B
	}
	\startdata
	\object{$\omega_m~(10^9~{\rm rad~s^{-1}})$} & 1.67 & 0.69 & ... & ... \\
	\object{$N_{\rm b}$} & 2 &  5 & $4\pm1$ &  $4\pm2$  \\
	\object{$\gamma$} & 214 &  $10^3$ & $214\pm15$ & $(4.9\pm2.2)\times10^3$  \\
	\object{$\log_{10}(\rho/{\rm cm})$}  & 6.5 &  6 & $6.5\pm0.1$ & $6.2\pm0.7$  \\
	\object{$\log_{10}N$}  &  46.45 &  42.85  &  $46.5\pm1.4$ & $45.8\pm0.9$  \\
	\object{$\Delta~({\rm cm})$}  &  2  &   9 & $3.5\pm0.5$  & $2.0\pm1.5$  \\
	\object{$\omega_{m,c}~(10^9~{\rm rad~s^{-1}})$} & ... &  ... & $1.68\pm0.05$ & $2.0\pm0.1$ \\
	\object{$\sigma_{\rm s}~(10^9~{\rm rad~s^{-1}})$} & ...  &  ... & $0.05\pm0.01$ & $0.14\pm0.02$ \\
	\object{$p_{\rm KS}$} & ...  &  ... & $<10^{-2}$ & 0.8 \\
	\enddata
\end{deluxetable*}

\subsection{Steep, Shallow, Narrow, and Broadband Spectral Analysis}
\label{subsec:steep}
As mentioned above, the period of the bunch distribution, represented by $1/\omega_m$, determines the interval between two adjacent peak frequencies. For a given $\omega_m$, the number of bunches per cluster, $N_{\rm b}$, determines the slenderness of the spectrum at each peak frequency, as shown in the top panel of Figure \ref{fig:steep}. In this plot, we vary $N_{\rm b}$ from 2 to 50, while keeping the other parameters constant: $\rho=10^6~{\rm cm}$, $\gamma=10^3$, $\omega_m=1.68\times10^9~{\rm rad~s^{-1}}$, $N=10^{45}$, $\Delta=5~{\rm cm}$, and $D=974.4~{\rm Mpc}$ for a redshift of $z=0.1927$. 
	
Let's focus on the spectrum along the first peak frequency. If the spectra for $N_{\rm b}=2$, 3, 5, 10, and 50 exhibit comparable peak fluxes by only adjusting $N$ from $2.25\times10^{45}$ to $3.6\times10^{42}$, as shown in the middle panel of Figure \ref{fig:steep}, it is clear that the observed spectrum above the telescope sensitivity becomes narrower and steeper as $N_{\rm b}$ increases. 
A more detail analysis is as follows. In the middle panel of Figure \ref{fig:steep}, the Gaussian-like spectra above the telescope sensitivity can be approximately fitted with a symmetric broken power-law function 
\begin{equation}
	F_\nu=F_{\nu, 0} \begin{cases}\left(\frac{\nu}{\nu_0}\right)^{\alpha}, & \text { if } \nu \leqslant \nu_0, \\ \left(\frac{\nu}{\nu_0}\right)^{-\alpha}, & \text { if } \nu>\nu_0.\end{cases}
	\label{eq:broken}
\end{equation}
For further simplicity, although somewhat roughly, 
one can use the slopes $\alpha$ of the dashed-dotted lines to represent the power-law fitting results for the falling half part of the spectra.
In this case, one can easily obtain power-law indices $\alpha=0.8$, 6, 12, 25, and 128, in correspondence to relative spectral bandwidth $\Delta\nu/\nu_0=0.46$, 0.46, 0.27, 0.13, and 0.02, for $N_{\rm b}=2$, 3, 5, 10, and 50, respectively. Moreover, it is clear that the spectral index is inversely correlated with the bandwidth.

Based on above analysis, broadband spectra typically defined by $\Delta\nu/\nu_0>0.3$ appearing in most FRB bursts while narrow spectra only occurring in a few bursts indicates that the bunch number per cluster $N_{\rm b}$ is generally small, perhaps $\lesssim5$ for most bursts. This signifies that the formation of clusters composed of a large number of bunches is not easy. 
It is notable that this result also depends on the telescope sensitivity, 
in addition to parameters like $N_{\rm b}$, $N$, $\rho$, and $\Delta$. 
As illustrated in the middle panel of Figure \ref{fig:steep}, if the value of the telescope sensitivity is lower and a spectrum like the one of $N_{\rm b}=5$ is given, the observed spectrum would appear broader. Furthermore, if the telescope's band envelope only covers the rising or falling half of the spectrum, the observed spectrum may exhibit a single power-law shape rather than a complete Gaussian-like shape or broken power-law shape.

Let's use previous analysis to conduct a case study for the most fractionally narrow-banded burst in FRB 20190711A detected by the Ultra-Wideband Low(UWL) receiver system \citep{kumar21}. As shown in the bottom panel of Figure \ref{fig:steep}, the observed narrow-banded spectrum can be well fit using a quasi-periodic spectrum with the following parameter values $N_{\rm b}=25$, $\rho=10^7~{\rm cm}$, $\gamma=10^2$, $\omega_m=2.06\times10^9~{\rm rad~s^{-1}}$, $N=5.3\times10^{46}$, $\Delta=5~{\rm cm}$, and $D=3081.7~{\rm Mpc}$ for a redshift of $z=0.522$. One can see that this observed extremely narrow spectrum is relevant to a relatively large $N_{\rm b}$. Moreover, one can also see that the quasi-periodic spectrum can also explain why \cite{kumar21} found no evidence of any emission in the remaining part of the 3.3 GHz UWL band, as its flux around $\sim3.3$ GHz is lower than the instrument sensitivity.

\begin{figure}
	\includegraphics[scale=0.55]{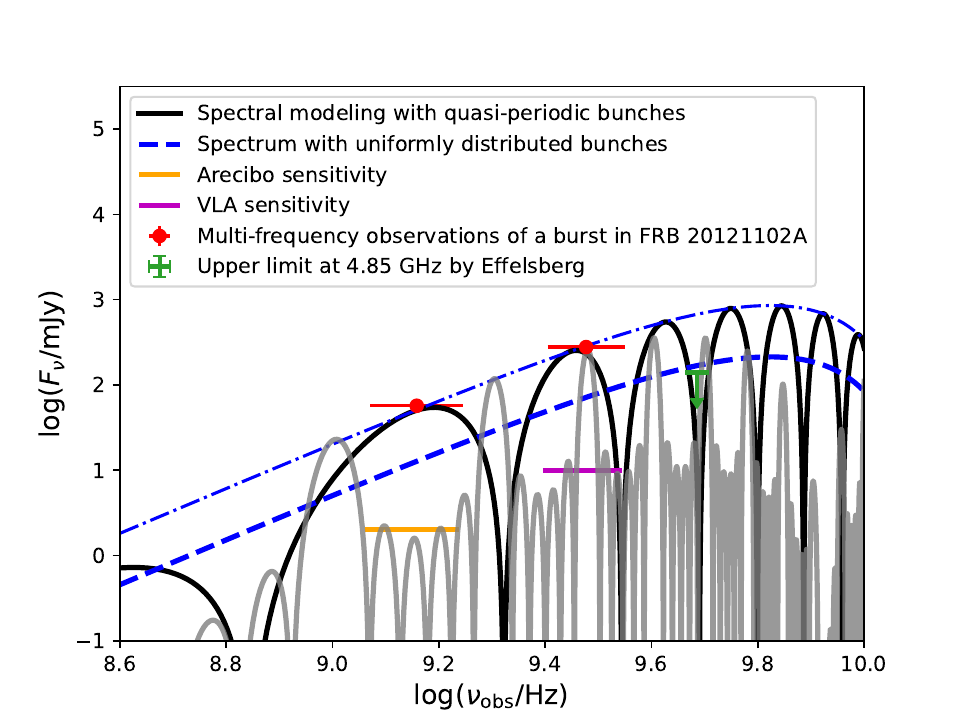}
	\includegraphics[scale=0.55]{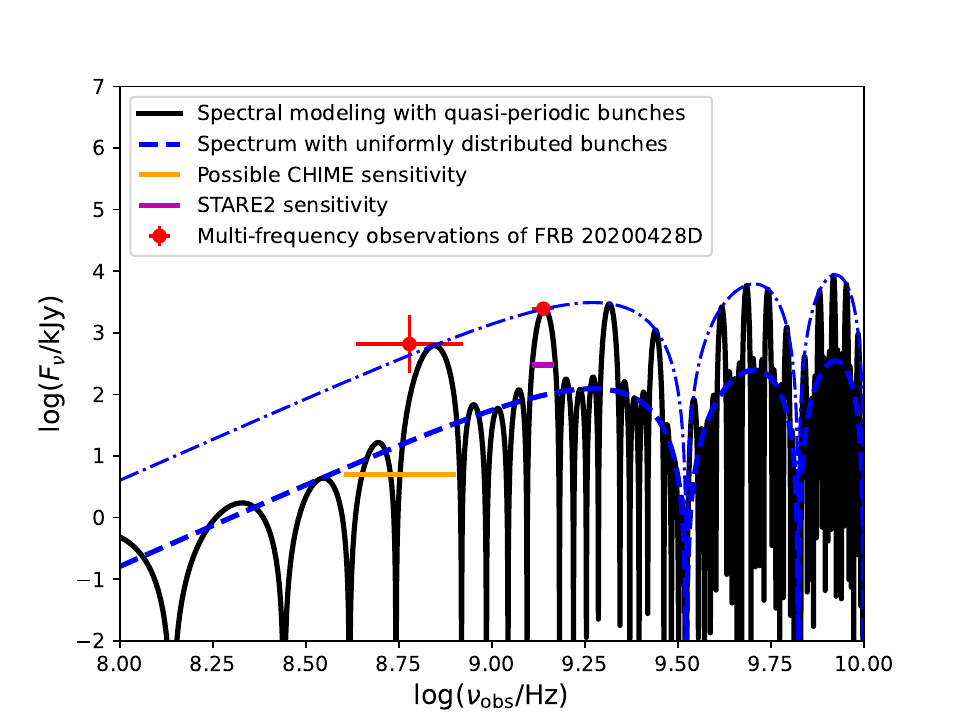}
	\caption{The multi-frequency simultaneous spectra (red points) for FRBs 20121102A (upper panel) and 20200428D (lower panel) are modeled using coherent curvature radiation from quasi-periodically structured bunches (black solid lines). For comparison, spectra generated from uniformly distributed bunches (blue dashed lines) are also shown. The dashed-dotted lines represent vertically shifted versions of the spectra for uniformly distributed bunches. It is noteworthy that the flux values at the coherent peak frequencies of the spectra with quasi-periodically structured bunches lie precisely on these dashed-dotted lines. Additionally, the non-detection at $4.85\pm0.25$ GHz, conducted simultaneously by Effelsberg telescope, is indicated by a green upper limit of $\sim140$ mJy. The gray line in the upper panel is used to illustrate the possibility for explaining the non-detections from the Arecibo despite being detected at the VLA for several bursts in \cite{law17}.}
	\label{fig:multifreq}
\end{figure}

\subsection{Modeling Multi-frequency Simultaneous Spectra}
\label{subsec:multifrequency}
The multi-frequency simultaneous spectra for the FRB 20121102A burst, along with the telescope sensitivities, were sourced from \cite{law17}. For FRB 20200428D, we obtain the spectral data and sensitivities from the Survey for Transient Astronomical Radio Emission 2 (STARE2) instrument \citep{boc20} and the CHIME instrument for the simultaneous second component \citep{chime20}. Note that all flux and fluence measurements are subject to a systematic uncertainty of roughly a factor of 2 for the CHIME data related to FRB 20200428D \citep{chime20}. Since the exact sensitivity of CHIME for FRB 20200428D is not available, we pick one upper limit from the CHIME/FRB non-detections of November 2019 high-energy bursts from SGR 1935+2154 \citep{chime20} as a possible estimate, although this approach is not rigorous.

We model these multi-frequency simultaneous spectra using Equations (\ref{eq:dI_tot})-(\ref{eq:Fv}). The values of parameters, such as the distribution property of quasi-periodic structured bunches $\omega_m$, the bunch number in a cluster $N_{\rm b}$, the Lorentz factor $\gamma$, the curvature radius $\rho$, 
the separation between electron and positron clumps $\Delta$, and $N$, are listed in Table \ref{tab:modeling}. From the modeling results in Figure \ref{fig:multifreq}, one can see that the different observed frequencies may exactly cover $\omega_{\rm obs}=\omega /(1+z) =2 n \pi \omega_m /(1+z)$ for different values of $n$. In this case, the non-detection at $4.85\pm0.25$ GHz conducted simultaneously by Effelsberg telescope\footnote{Given the power-law index $\alpha>1.4$ for $F_{\nu}\propto \nu^{-\alpha}$, derived from the observations at 3 GHz by GBT C-band and 4.85 GHz by Effelsberg \citep{law17}, the upper limit of the flux at 4.85 GHz is $\sim140$ mJy.} 
for the FRB 20121102A burst is obvious, as displayed in the upper panel of Figure \ref{fig:multifreq}. 
From the results in Table \ref{tab:modeling}, it is evident that the bunch number in a cluster $N_{\rm b}$ is likewise relatively small for both FRBs 20121102A and 20200428D. 
	
Additionally, in either the upper or lower panel of Figure \ref{fig:multifreq}, the flux values at coherent peak frequencies $2n\pi\omega_m$ align precisely with the dashed-dotted line denoting a vertical shift of the spectrum for uniformly distributed bunches. 
This suggests that the flux variation at different coherent peak frequencies follows the spectral shape characteristic of uniformly distributed bunches. 
Regarding the slope in logspace between the two spectral data points in either FRB 20121102A or FRB 20200428D, it is significantly steeper than 2/3, which cannot be explained by the rising $F_{\nu}\propto \nu^{2/3}$ spectral shape from commonly-discussed bunches described in \cite{yang18}. Nevertheless, this relatively steep slope can be accounted for by the spectrum of separated electron-positron pair bunches, which exhibits a rising slope of 8/3 \citep{yang20}. This is why we employed separated electron-positron pair bunches in Equation (\ref{eq:E_N}).  
It should be noted that \cite{yang20} specifically investigated the spectrum related to separated electron-positron pair bunches to interpret the relatively steep slope observed between the two spectral data points in FRB 20200428D. While in our study, we explain it using the spectrum from quasi-periodic structured bunches composed of separated electron and positron clumps.

\begin{figure}
	\includegraphics[scale=0.55]{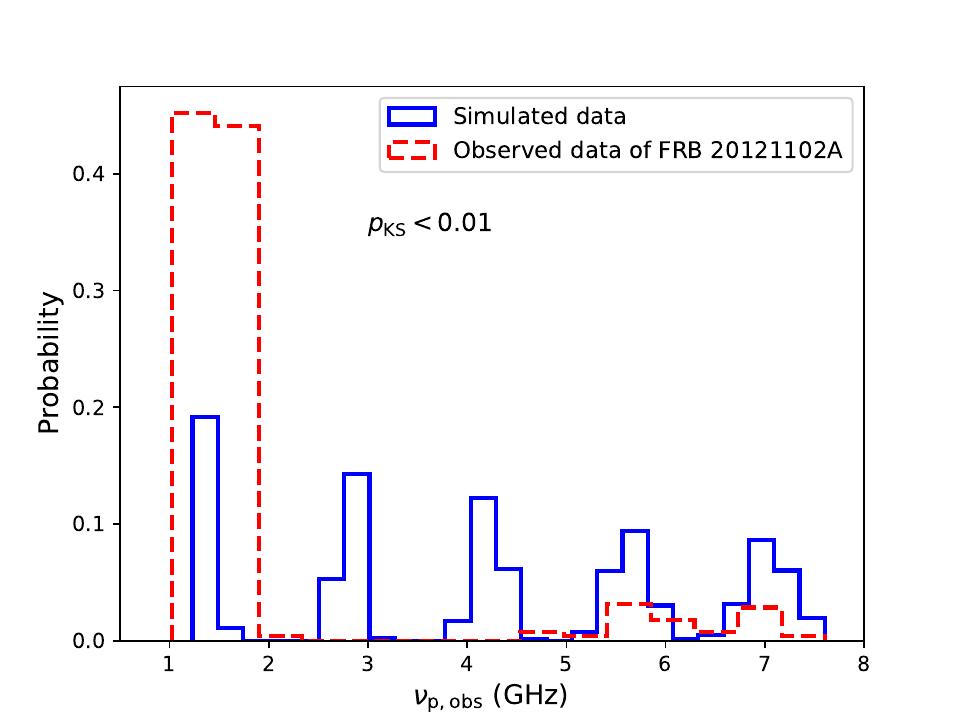}
	\includegraphics[scale=0.55]{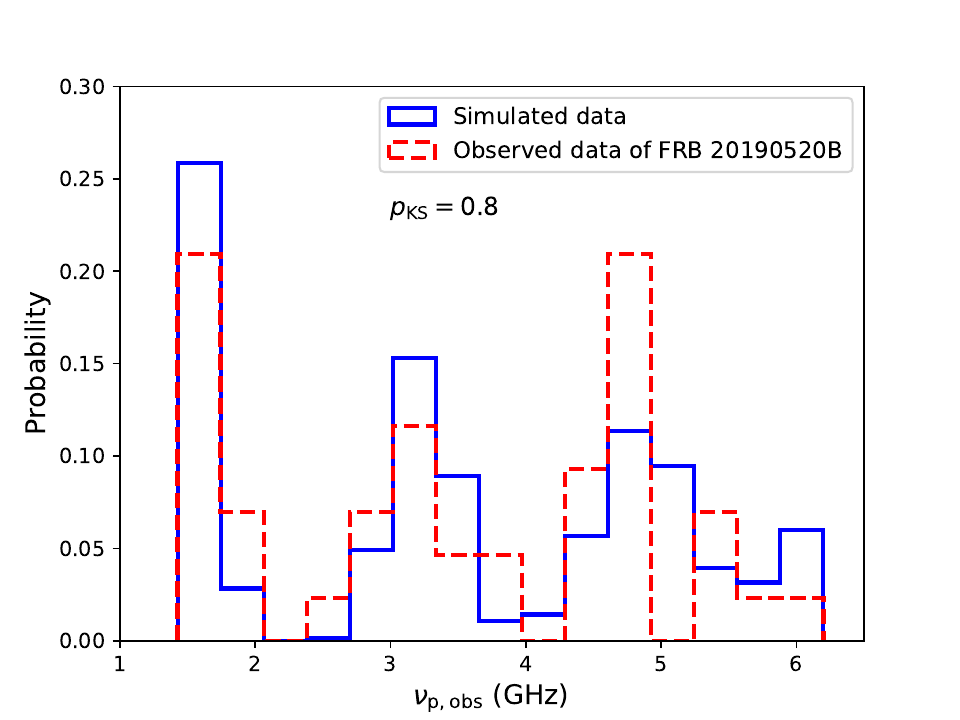}
	\caption{Simulating the fringe patterns of peak frequency distributions observed in the bursts of FRB 20121102A and FRB 20190520B, based on the assumption that the bursts exhibit an intrinsic quasi-periodic spectrum produced by coherent curvature radiation from quasi-periodic structured bunches.}
	\label{fig:fringe}
\end{figure}

\subsection{Simulating Statistical Fringe Spectra}
\label{subsec:fringe}
The fringe patterns of peak frequency statistical distributions in FRBs 20121102A and 20190520B were identified by \cite{lyu22} and \cite{lyu23}, respectively. 
For FRB 20121102A, they used data from \cite{zhangyg18} for GBT C-band observations \footnote{The data from \cite{zhangyg18} contained those already reported in \cite{gaj18}.} 
and \cite{hew22} for Arecibo observations. In this study, we use the data from \cite{zhangyg18} and \cite{jah23} because the data from \cite{hew22} do not provide a well-determined central frequency. For FRB 20190520B, we use data collected by \cite{lyu23} from \cite{niu22}, \cite{feng22}, \cite{dai22}, and \cite{anna23}.
Doing MC simulations using the same method as in \cite{xie20}, we outline the steps as follows:
\begin{itemize}
\item First, we assume each burst has an intrinsic quasi-periodic spectrum with a period of $1/\omega_m$ based on Equations (\ref{eq:dI_tot})-(\ref{eq:Fv}). 
Moreover, we assume that the $\omega_m$ values of the bursts in a given FRB are quasi-universal, following a Gaussian distribution as
\begin{equation}
	\Phi(\omega_m)\propto\frac{1}{\sqrt{2 \pi}\sigma_{\rm s} } \exp \left[\frac{-\left(\omega_m-\omega_{m,c}\right)^2}{2 \sigma_{\rm s}^2}\right],
	\label{eq:Phi}
\end{equation}
where $\omega_{m,c}$ is the expected value and $\sigma_{\rm s}$ is the standard derivation. Although coherent peak frequencies are determined by $\omega=2n\pi \omega_m$, a complete quasi-periodic spectrum for a burst also depends on parameters $N_{\rm b}$, $\gamma$, $\rho$, $N$, and $\Delta$, as they affect the spectral profile. For each parameter set $\left\{\omega_{m,c}, \sigma_{\rm s}, N_{\rm b}, \gamma, \rho, N, \Delta\right\}$, 
we randomly generate an $\omega_m$ value to simulate a complete quasi-periodic spectrum for a burst via a bootstrap method based on Equation (\ref{eq:Phi}). 
Because the flux peaks at coherent peak frequencies $\nu_{\rm p}=n\omega_m$\footnote{The peak frequencies $\nu_{\rm p}=\omega_{\rm p}/(2\pi)=2n\pi\omega_m/(2\pi)=n\omega_m$.}, in real observations, one can assume that the observation covers only one of $n\omega_m$, with the integer $n$ treated as a random number in the range of $n\in I(1,10)$. From the generated spectrum, one then can obtain the flux at $F_{\nu}(\nu_{\rm p})$, where $\nu_{\rm p}$ relates to its observer-frame frequency via $\nu_{\rm p, obs}=\nu_{\rm p}/(1+z)$. 
Further we compare this flux $F_{\nu}(\nu_{\rm p,obs})$ with the telescope sensitivity $F_{\min}$ at corresponding frequency band. Only if $F_{\min}<F_{\nu}<10^5F_{\min}$\footnote{We use the limit $F_{\nu}<10^5F_{\min}$ as there is no peak flux exceeding this threshold in the observed bursts of FRBs 20121102A and 20190520B.}, we retain this flux $F_{\nu}(\nu_{\rm p,obs})$ and record its associated $\nu_{\rm p,obs}$. In this way, we obtain one or zero $\nu_{\rm p,obs}$ for one burst.
\item Second, for each set of $\left\{\omega_{m,c}, \sigma_{\rm s}, N_{\rm b}, \gamma, \rho, N, \Delta\right\}$, we randomly generate $10^3$ $\omega_m$ values\footnote{This number $10^3$ is chosen to be close to the number of observed bursts for FRBs 20121102A and 20190520B.} to simulate $10^3$ quasi-periodic spectra for $10^3$ bursts, resulting in $\leqslant10^3$ $\nu_{\rm p,obs}$ values. 
We then compare this simulated $\nu_{\rm p,obs}$ sample with the observed sample using the Kolmogorov–Smirnov (K-S) test.  
\item Third, since each simulated sample corresponds to a set of $\left\{\omega_{m,c}, \sigma_{\rm s}, N_{\rm b}, \gamma, \rho, N, \Delta\right\}$, we generate $10^4$ such sets to produce $10^4$ simulated samples. 
The parameter values are randomly drawn from uniform distributions $(\omega_{m,c}/{10^9 \rm rad~s^{-1}}) \in U(0.1,5)$, 
$(\sigma_{\rm s}/{10^9 \rm rad~s^{-1}})\in U(0.01,0.5)$,
$N_{\rm b}\in I(2,50)$, $\log_{10}\gamma\in U(2,4)$, $\log_{10}(\rho/{\rm cm})\in U(5,7)$, $\log_{10}N\in U(40,50)$, 
and $(\Delta/{\rm cm})\in U(1,20)$. 
The prior distributions for $\omega_{m,c}$, $\sigma_{\rm s}$, $N_{\rm b}$, $N$, and $\Delta$ 
are based on our results in Sections \ref{subsec:steep} and \ref{subsec:multifrequency}.
While the priors of $\gamma$ and $\rho$ are typical for coherent curvature radiation with critical frequency $\nu_c=\omega_c/(2\pi)\sim{\rm GHz}$ within the magnetosphere \citep[e.g.,][]{kumar17,yang18}.
By comparing the simulated samples with the observed sample, we select the best-fitting sample based on the highest K-S test value $p_{\rm KS}$, as displayed in Figure \ref{fig:fringe} for FRBs 20121102A and 20190520B. The corresponding parameter values and their $1\sigma$ confidence levels are listed in Table \ref{tab:modeling}. The confidence levels are obtained by plotting K-S test value contours in the parameter planes, as demonstrated in the figure 1 of \cite{xie20}.
\end{itemize}

\subsubsection{Results and Implications}
\label{subsubsec:results}
As shown in the results, the simulated data for FRB 20190520B exhibit a high degree of consistency with the observed data, achieving a maximum $p$-value of $p_{\rm KS}=0.8$. For FRB 20121102A, the main peaks in the simulated data align with the observed data, though the $p$-value is low. The low $p$-value may be attributed to the incompleteness of the observed sample because of the lack of data from 2 GHz to 5 GHz and other observational effects, 
such as differences in observation periods and sensitivity thresholds among various telescopes. 
Additionally, the $\omega_m$ values for bursts within a single FRB show significant universality. For example, $(\sigma_{\rm s}/10^9~{\rm rad~s^{-1}})=0.05$ and 0.14 for FRBs 20121102A and 20190520B, respectively. This may imply a consistent bunch structure possibly relying on the properties of the FRB source such as a neutron star (NS). 

If FRBs are generated through coherent curvature radiation, 
the structured bunches might be formed through pair cascades in charge-starvation regions or via two-stream instability \citep{wang24}. 
If the structured bunches result from sudden pair cascades in a charge-starvation region, similar to the continuous sparking from the polar cap in regular radio pulsars, the periodicity of these cascades is estimated by \citep{tim15}
\begin{equation}
P_{\rm cas} \sim 3h_{\rm gap}/c \sim 1 \times 10^{-6} \rho_7^{2 / 7} P^{3 / 7} B_{12}^{-4 / 7}|\cos \beta|^{-3 / 7}\  {\rm s},
\label{eq:P_cas}
\end{equation}
where $\rho_7=\rho/10^7{\rm cm}$, $P$ is the spin period, $B_{12}=B/10^{12}{\rm G}$ is the surface dipole magnetic field, $\beta$ is the inclination angle of the magnetic axis.
For a single NS, its $P$, $B$, and $\beta$ remain nearly constant over a long time. The highly universal $\omega_m=1/P_{\rm cas}$ values\footnote{Note that $P_m=1/\omega_m$ is the period of the bunch distribution.} centered around $\omega_{m,c}$, obtained from the simulations, suggest universal $\rho$ values for the bursts within an FRB. 
From the derived curvature radii, $\log_{10}(\rho/{\rm cm})=6.5$ for FRB 20121102A and 6.2 for FRB 20190520B, one could constrain the spin period and surface dipole field of their associated NSs. For instance, assuming a spin period $\sim1$ s and inclination angle $\beta=0$, 
the surface dipole fields for both NSs are as high as $10^{17}$ G, supporting the magnetar origin of FRBs.

Alternatively, if the structured bunches are generated via two-stream instability which invokes Langmuir wave for resonant reactive instability as the bunching mechanism, the bunch separation would be related to the wavelength of the Langmuir wave at the breakdown of the linear regime, i.e., 
\begin{equation}
	l_{\text {sep}} \sim \frac{2\pi c}{\omega_{\rm L}}=cP_m,
\end{equation}
where $\omega_{\rm L}=2\gamma_{j=2}^{1 / 2} \omega_{\mathrm{p}, j=2}$ is the characteristic frequency of the Langmuir wave, with $\gamma_{j=2}$ and $\omega_{\mathrm{p}, j=2}$ being the Lorentz factor and plasma frequency of the plasma component two, respectively. For more details, see \cite{wang24}. 
According to the simulation results in Table \ref{tab:modeling},  
the characteristic frequencies of the Langmuir waves are $\nu_{\rm L}=\omega_{\rm L}/(2\pi)=\omega_{m,c}\sim1.68\times10^9$ Hz 
and $\sim2.0\times10^9$ Hz for FRBs 20121102A and 20190520B, respectively.

\section{Summary and Discussion}
\label{sec:summary}
We have demonstrated that the observed steep and shallow spectra, narrow and broadband spectra, multi-frequency simultaneous spectra, and statistical fringe spectra in FRBs may be various manifestations of intrinsic quasi-periodic spectra, arising from coherent curvature radiation by structured bunches composed of separated electron and positron clumps. 
Our analysis, modelings, and MC simulations lead to the following main results:
\begin{itemize}
	\item For a given bunch distribution period $\omega_m$, the slenderness of the spectrum at each peak frequency is determined by the number of bunches per cluster, $N_{\rm b}$. In general, the observed spectrum above the telescope sensitivity becomes narrower and steeper as $N_{\rm b}$ increases. Moreover, the spectral index is inversely correlated with the bandwidth. The appearance of broadband spectra in most FRB bursts indicates that the number of bunches per cluster is typical small, perhaps $N_{\rm b}\lesssim5$ for most bursts, although this is also dependent on the telescope sensitivity. Furthermore, if the telescope's band envelope only covers the rising or falling half of the spectrum, the observed spectrum may exhibit a single power-law shape rather than a full Gaussian-like or broken power-law profile. In addition, a case study for the most fractionally narrow-banded burst in FRB 20190711A was conducted. Its extremely narrow spectrum can be well fit by a quasi-periodic spectrum relevant to a relatively large $N_{\rm b}=25$.
	\item The observed multi-frequency simultaneous spectra for FRB 20200428D and a burst in FRB 20121102A can be effectively modeled by intrinsic quasi-periodic spectra within typical parameter ranges, as illustrated in Figure \ref{fig:multifreq}. From the results in Table \ref{tab:modeling}, 
	it is evident that the number of bunches per cluster, $N_{\rm b}$, is likewise relatively small for both FRBs 20121102A and 20200428D. 
	\item The statistical fringe patterns in peak frequency distributions for FRBs 20121102A and 20190520B can be reproduced if their bursts display intrinsic quasi-periodic spectra with highly universal $\omega_m$ values. Additionally, we discuss potential formation mechanisms of structured bunches. If these bunches originate from pair cascades in charge-starvation regions, the period of the bunch distribution reflects the period of pair cascades, which could be used to constrain the properties of the FRB source, such as its magnetic field. Alternatively, if the bunches are formed via two-stream instability involving Langmuir wave, the period of the bunch distribution could help estimate the wavelength of the Langmuir wave at the point where the linear regime breaks down.
\end{itemize}	

As to whether the quasi-periodic bunch model can explain burst-to-burst variation in the spectra for an FRB, we illustrate as follows. FRB 20121102A shows significant spectral variation from burst to burst. For example, although one burst was detected simultaneously by both the Arecibo and VLA, several bursts in \cite{law17} were detected only by the VLA, with no detections by the Arecibo. If the distribution properties of quasi-periodic structured bunches, specifically, the $\omega_m$ values are not centered around a single universal $\omega_{m,c}$ value, then the quasi-periodic bunch model not only can explain the multi-frequency simultaneous spectra for the FRB 20121102A burst (as shown by the black solid line in the upper panel of Figure \ref{fig:multifreq}, under the parameter values in Table \ref{tab:modeling}), but also possibly explain the non-detections from the Arecibo despite being detected at the VLA for several bursts (shown by the gray solid line in the same panel of Figure \ref{fig:multifreq}, under the parameter values $\omega_{m,c}=1.2\times10^9~{\rm rad~s^{-1}}$, $N_{\rm b}=6$, $\gamma=214$, $\log_{10}(\rho/{\rm cm})=6.5$, $\log_{10}N=45.2$, and $\Delta=3$). Although this requires two distinct $\omega_{m,c}$ values, such a requirement is also noted in the fit in the upper left panel of Figure 1 of \cite{lyu22}. This may be another reason why a relatively large $p_{\rm KS}$ value cannot be achieved when simulating with a single universal $\omega_{m,c}$ value for the fringe pattern of peak frequency distribution in FRB 20121102A bursts, in addition to the incompleteness of the observed sample and other observational effects mentioned in Section \ref{subsubsec:results}.

From a prospective viewpoint, we think that the intrinsic quasi-periodic spectrum may also account for the spectral zebra pattern observed in the high-frequency interpulse of the Crab pulsar’s radio emission \citep{han07,han16}. While several models have been suggested to explain this phenomenon \citep[see review,][]{eilek16}, including the recent magnetosphere diffraction screen model \citep{med24}, the quasi-periodic spectrum could provide a compelling alternative explanation. This will be explored elsewhere.

\acknowledgments
We are very grateful to the referee for careful and thoughtful comments and suggestions that have helped improve this manuscript substantially.
This work is supported by the starting Foundation of Guangxi University of Science and Technology (grant No. 24Z17).


\end{document}